# FACILITY BASED ON ELECTRON LINAC LUE-40 - CURRENT STATE AND MAIN RESEARCH DIRECTIONS

M.I. Ayzatsky[1], V.M Boriskin[1], I. O. Chertishchev[1], O.S. Deiev[1], K.Yu. Kramarenko[1], V.A. Kushnir[1], V.V. Mytrochenko[1,3], S.M. Olejnik[1], S.O. Perezhogin[1], S.M. Potin[1] L.I. Selivanov[1], I.S. Timchenko[1,2], Yu. O. Titarenko[1], V.Yu. Tytov[1], V.L. Uvarov[1]

[1] NSC "Kharkiv Institute of Physics and Technology", National Academy of Sciences of Ukraine, 1 Academichna Str., Kharkiv 61108, Ukraine
[2] Institute of Physics, Slovak Academy of Sciences, SK-84511 Bratislava, Slovakia
[3] CNRS/IJCLAB, 15 Rue Georges Clemenceau Str., Orsay, 91400, France
E-mail: vakdiver@gmail.com

The paper describes the state of the installation based on the linear accelerator LUE-40 after restoration in 2023. Additional devices have been developed and created to conduct nuclear physics research. The main beam parameters are given. The electron energy range has been expanded and is currently 16–95 MeV with an average beam current of up to 6 µA. The main areas of scientific research are described and the results are presented.

## INTRODUCTION

The S-band electron linac LU-40 was created during the construction of the LU-2 accelerator with energy of 2 GeV. Initially, LU-40 was intended for testing accelerating sections before their installation in LU-2. However, after 1965, when a large accelerator was put into operation, the LU-40 began to be used for scientific research. The accelerator consisted of a diode gun, an injector section and one homogeneous accelerating section of the "Kharkov-65" type. The energy at the accelerator output was 40 MeV and the pulse current was about 1 A. Among the experimental research conducted in the period 1970 – 2000, it is worth noting: research on the interaction of relativistic electrons with crystals, including the study of parametric radiation, study of particle acceleration processes in plasma, study of the electrodynamic characteristics of new type accelerating sections, study of the influence of electron and gamma radiation on the properties of various materials

In connection with the need to conduct a new level of scientific and applied research, a significant reconstruction of the accelerator was carried out in 2004 - a new injector and accelerating sections were installed, the control systems and beam parameter measurements were modernized. In fact, a new accelerator called LUE-40 was created. The maximum energy of the particles at the output reached 100 MeV. The maximum average current of 6 µA was limited by the existing radiation protection. After reconstruction, research was conducted in several main directions: study of photonuclear methods for obtaining isotopes for medical purposes, study of photonuclear reactions with multiple neutron yields, study of the effect of radiation on the properties of dielectrics, study of ballistic electron bunching, and radiation testing of materials for high-energy physics detectors. Unfortunately, because of military operations in 2022, almost all the accelerator systems were damaged. Some of them were completely disabled. Repair work and adjustment of the accelerator were carried out by the employees of the "Accelerator" Science and Research Establishment and lasted for 10 months. At present, the LUE-40 has been fully restored and the beam parameters correspond to the passport values, which made it possible to continue experimental research.

## 1. LINEAR ACCELERATOR

The LUE-40 linac consists mainly of the 25kV DC electron gun, the buncher and two 4 m long accelerating sections of the "Kharkov-85" type [1]. A simplified diagram of the accelerator is shown in Fig.1. The electron gun is directly connected to the buncher on non-propagating oscillations [2]. The bunching system consists of 5 coupled resonators, in which, due to a special choice of cell sizes, an exponentially increasing longitudinal electric field is realized [3]. A bunching system allows one to effectively form particle bunches from a continuous beam emitted from the cathode of such low-voltage guns and to accelerate particles to 700-900 keV. To ensure normal operation of the buncher, a power of 750 kW is supplied to its resonance system. Microwave power is transmitted to the resonant system through waveguide-coaxial transition. The use of coaxial microwave power input allows to eliminate RF field asymmetry and improve beam emittance.

The accelerating section of the "Kharkov-85" type consists of four subsections with different shunt impedance. This allows maintaining a quasi-constant field strength along the section. These accelerating sections are distinguished by a significant acceleration rate and a low group velocity (average value of 0.016 c). The microwave power supply of the LUE-40 is carried out from two klystron amplifiers with an output power of up to 18 MW, the pulse duration of 2.50 µs and the repetition rate of 50 Hz. The beam energy reaches 100 MeV with a microwave power supply of about 13 MW in each section.

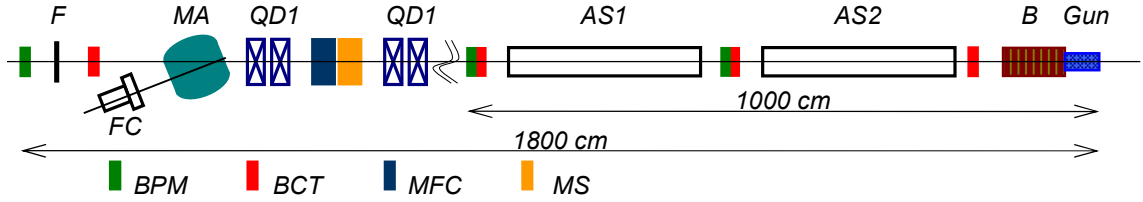

*Fig.1 Schematic layout of the LUE-40 linac. B-buncher, AS-accelerating section, QD - quadrupole doublet, MFC – moving Faraday cup, MS – moving slots, BPM - beam position monitor, BCT – beam current transformer, MA- magnetic energy analyzer, FC - slit + Faraday cup, F – outlet foil*

Beam measurement system includes devices that can provide information about the beam intensity, and its change during the pulse, beam position, energy, energy spectrum and spatial distribution of the beam density. The accelerator is equipped with four beam current transformers (BCT), Faraday cup and three beam position monitors (BPM). One of the BMPs is installed directly behind the outlet foil [4]. Beam size is measured using both movable slit collimators and a wire scanner. The energy and energy spectrum of the beam are measured using a magnetic analyzer with energy resolution of 0.2 %.

Until recently, the range of beam energy change at the accelerator output was 35-95 MeV. For certain tasks, such as studying photonuclear reactions in the giant dipole resonance range (GDR), the electron energy must be 20-30 MeV. Unfortunately, the most obvious solution to this problem – a sufficient reducing the field amplitude in the first section - cannot be implemented with the existing microwave power supply scheme without reducing the field amplitude in the buncher. It is quite a difficult task to obtain at the linac output the beam with an energy of less than 35 MeV while maintaining the spectral and spatial characteristics. The solution to this problem required both numerical simulation of the self-consistent dynamics of particles in the accelerator and analysis of the klystron operation in non-standard modes (output power 1-2 MW). Based on the numerical simulation, the amplitude and phase of the "braking" field in the second section of the accelerator were determined. In experiments, a beam with an energy of 16 MeV and an energy spread less than 4 % for 70 % of particles was obtained at the accelerator output, which may be acceptable for some experiments.

Main parameters of the beam after the recovery of the accelerator in 2023 are shown in Table.1. The values of energy spread and beam sizes given in the table are integral during the current pulse 1.6 μs. As measurements have shown, they significantly exceed the corresponding values measured at a given moment (time frame 50 ns).

The accelerator can be adjusted to any electron energy in the range of 16..95 MeV by changing the power and phase of the RF supply only of the second section. Parameters of the microwave power supply of the injectors and the first section remain unchanged., which made it possible to maintain the same conditions for the bunch formations. Typical beam energy spectra are shown in Fig.2 .

Table 1

| Parameter | Value |
|---|---|
| RF frequency [MHz] | 2797.09 |
| Beam energy [MeV] | 16…95 |
| Energy spread (70% of beam) [%] | < 2.5 |
| Beam pulse current [mA] | up to 70 |
| Average beam current [μA] | up to 6 |
| Normalized emittance [mm·mrad] | < 60 |
| Beam size, FWHM [mm]* | < 6 |
| Beam pulse duration [μs] | 1.65 |
| Repetition rate [Hz] | 50 |

*measured at a distance of 6.5m from second section exit with the focusing system off*

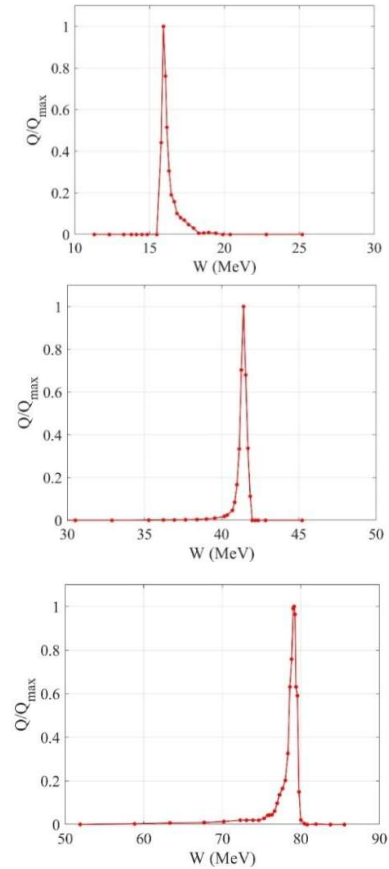

*Fig.2*

It is known that when electrons are accelerated in the accelerating section, electron bunches excite a field in the structure that interacts with the electrons. As a result, this leads to a decrease in the beam energy. This phenomenon, called "beam loading", significantly affects the beam parameters during acceleration in sections with high shunt impedance. The measured dependence of the average beam energy on the beam current is shown in Fig.3. From the measurement results it follows that the "beam loading" value is 63 MeV/A per section, which corresponds to the simulation results.

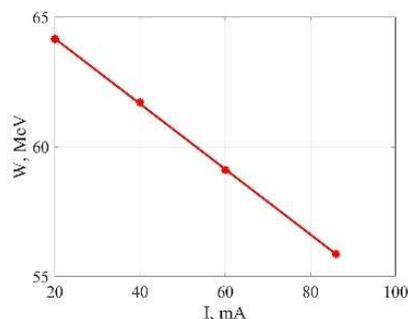

*Fig.3 Average beam energy vs. the beam pulse current*

## 2. OUTPUT DEVICES

To conduct experiments at the accelerator output, various devices are used to form the necessary fields and radiation intensity with both primary electrons and bremsstrahlung gamma quanta or neutrons. To generate bremsstrahlung radiation, converters with a thickness of 0.1 to 4 mm with air or water cooling have been developed and are used in accordance with the experimental program. Several types of converter devices for irradiating samples with mixed neutron and gamma radiation have been developed and created [7,8]. In particular, the converter for mainly neutron irradiation of elements of high-energy physics detectors consists of 10 tungsten plates with 1 mm spaces between them in order to air cooling effectively and an additional lead shell. According to numerical simulation the source makes it possible to obtain the neutron $4\pi$ flux of $3.6 \cdot 10^{11}$ s$^{-1}$ at electron energy of 85 MeV and an average electron beam intensity 5 μA ($3.12 \cdot 10^{13}$ s$^{-1}$). The most probable and average energies of neutrons are 0.7 MeV and 2 MeV, respectively.

When studying high-threshold photonuclear reactions, it is extremely undesirable for electrons passing through the converter to hit the sample being studied. Installing an electron absorber (for example, Al) between the converter and the sample basically solves this problem but also creates additional difficulties. The disadvantages of this scheme are the distortion of the shape of the bremsstrahlung spectrum, and the additional generation of photoneutrons, which also contribute to studied reaction yield. In order to divert the electrons that have passed through the converter and to obtain a "pure" beam of bremsstrahlung gamma quanta at the target, bending system was designed, manufactured and installed at the linac output [9]. The system is based on permanent magnets of rectangular cross-sections. The maximum on-axis field is 0.9 T, which provides sufficient separation of the electron beam and gamma rays at more than 90 mm from the magnet. The magnet, used to clean the gamma beam at an electron beam energy of up to 85 MeV and a converter thickness of up to 0.3 mm.

One of the LUE-40 tasks is the irradiation of various materials and devices with a high-energy electron beam in order to study their radiation resistance. In some cases, it is necessary to either ensure uniformity of the absorbed dose throughout the entire sample, or simultaneous identical irradiation of various samples located in a plane orthogonal to the beam axis. In this case the flatness of less than 10 % is most often required. To meet these requirements, a compact system consisting of scatterers with non-uniform thickness and a collimator is installed at the linac exit. This device allows one to obtain a fairly uniform 2D distribution of electron density and absorbed dose (Fig.4)

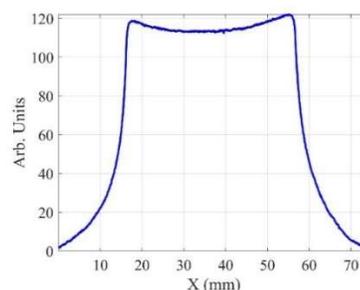

*Fig 4. The distribution of the electron density in the plane of the sample with a diameter of 20 mm. Beam energy is 42 MeV*

## 3. RESEARCH AT THE LUE-40 ACCELERATOR

### 3.1 RESEARCH IN APPLIED NUCLEAR PHYSICS

In recent decades, the studies were actively conducted on reactor-free methods of the medical isotope production, including the soft photonuclear technologies using bremsstrahlung radiation of an electron accelerator [10]. Such processes are being developed, in particular, to obtain the main diagnostic isotope $^{99m}$Tc, theranostic isotopes $^{67}$Cu and $^{47}$Sc, therapeutic nuclide $^{225}$Ac and others [11-14]. When evaluating the capacity and isotopic purity of such technologies, one of the challenges is the low accuracy of description of a number of photonuclear cross sections. For example, the spread of the available data on the cross section of the $^{48}$Ti(γ,p)$^{47}$Sc reaction, promising for the production of $^{47}$Sc, reaches 100% and still more (see, for example, [15,16]).

The facilities of the LUE-40 for the formation and diagnostics of an electron beam, deep regulation of its energy, as well as the creation of output devices for generating secondary bremsstrahlung and mixed X,n radiation provided the conditions for the development and experimental testing of new methods for determining cross sections of the photonuclear isotope production, as well as for the analysis and optimization of such processes.

So, a novel technique for estimating the amplitude and width of the giant dipole resonance was proposed [17]. Its testing on the reactions $^{100}$Mo($\gamma$,n)$^{99}$Mo and $^{58}$Ni($\gamma$,n)$^{57}$Ni has shown good agreement between the obtained and reference values [18]. By using this method at the LUE-40, the Lorentz-parameters of the photoproton cross-sections for titanium isotopes, including ones for the $^{48}$Ti($\gamma$,p)$^{47}$Sc reaction, were determined [19]. These data make it possible to estimate the yield of $^{47}$Sc and other scandium isotopes in a titanium-based target.

An analytical model was developed to describe and optimize the mode of the isotope production with the bremsstrahlung radiation in a thick technological target [20-22]. Based on the model, a new method for determining the photonuclear cross-section averaged over the bremsstrahlung photon flux was proposed [23]. Validation of the model and method was also carried out at the LUE-40.

The generation of isotopes in the field of mixed X,n radiation was studied also. For this, an output device was developed embracing a bremsstrahlung converter and a target located in the center of a neutron moderator. It has been shown, that when a target from natural molybdenum is used in such a device, the yield of $^{99}$Mo increases due to its generation simultaneously in the two $^{100}$Mo($\gamma$,n)$^{99}$Mo and $^{98}$Mo(n,$\gamma$)$^{99}$Mo channels [24]. The yield of isotopes in the targets made of natural palladium and rhenium under mixed X,n irradiation was studied as well [25].

Currently at the LUE-40, the studies are being carried out aiming at improving the accuracy of the cross-section data for photonuclear production of a number of medical isotopes and optimization of such processes using the developed techniques.

### 3.2 RESEARCH ON MULTIPARTICLE PHOTONUCLEAR REACTIONS

Today, research into multiparticle photonuclear reactions is highly relevant. Most of the data on cross-sections for photonuclear reactions are important for traditional studies of the Giant Dipole Resonance (GDR), mechanisms of its excitation and decay including competition between statistical and direct processes in decay channels, GDR configurational and isospin splitting, sum rule exhaustion, etc. Experimental cross sections for photonuclear reactions are also widely used in various applications, primarily in astrophysics, medicine, design of fast reactors and accelerator driven sub-critical systems.

Despite the significant amount of experimental data available in international databases, there is still an insufficiency of data for the cross-sections of many-particle photoneutron reactions. Mainly, data for reactions resulting in the release of up to 4 neutrons can be found. In addition, experimental data from different laboratories do not always agree with each other, even in the GDR range, for example, [26]. Data for photonuclear reactions involving charged particles in the output channel are mainly limited to the region of light nuclei and energies within the GDR range.

The LUE-40 beam with output devices has the necessary beam parameters for performing research on multiparticle photonuclear reactions with the emission of up to 10 nucleons from the nucleus. The experiments used the activation method and offline $\gamma$-spectrometric technique. A pneumatic transport system enables measurement of cross-sections for reactions producing nuclei with half-lives as short as 10 minutes.

The LUE-40 was used to study photonuclear reactions on nuclei such as $^{nat}$Mo, $^{nat}$Ni, $^{181}$Ta, $^{93}$Nb, $^{nat}$Cu, and $^{27}$Al in the range of end-point energies of bremsstrahlung gamma quanta $E_{\gamma max}$=35–95 MeV [27-37]. The experimental results were presented in the form of cross-sections $\langle\sigma(E_{\gamma max})\rangle$ averaged over the bremsstrahlung flux, cross-sections per equivalent photon $\langle\sigma(E_{\gamma max})\rangle_Q$ and isomeric ratios IR($E_{\gamma max}$) of nuclei-products. The results of the experiments are presented in the international EXFOR database. A comparison was made with available literature data and with theoretical estimates calculated using the $\sigma(E)$ cross-sections from the TALYS code [38] for different models of the LD level density and GSF photon strength functions.

The possibility of using the reactions $^{100}$Mo($\gamma$,n)$^{99}$Mo, $^{27}$Al($\gamma$,x)$^{24}$Na, $^{93}$Nb($\gamma$,n)$^{92m}$Nb, $^{93}$Nb($\gamma$,3n)$^{90}$Nb, and $^{181}$Ta($\gamma$,n)$^{180g}$Ta as bremsstrahlung monitors is considered. The advantages and disadvantages of using these monitor reactions are investigated for the energy range of 30 – 100 MeV.

Two different experimental setups were used in the studies. In the first one, an aluminum absorber was used to remove electrons from the bremsstrahlung $\gamma$-flux, in the second one - a deflecting magnet. A comparison of the averaged cross sections $\langle\sigma(E_{\gamma max})\rangle$ measured using two experimental setups was carried out [39], and good agreement between the obtained results was shown.

In the near future, it is planned to continue research on multi-particle photonuclear reactions on atomic nuclei of medium and heavy masses in the region of end-point bremsstrahlung energies of $\gamma$-quanta $E_{\gamma max}$ up to 100 MeV. Special attention will be paid to photonuclear reactions with protons and small clusters in the output channel. This is of particular interest since data on cross-sections and isomeric ratios for ($\gamma$,ypxn) reactions are very limited. At the same time, theoretical calculations for such reactions are more sensitive to the choice of model parameters.

Research on absolute cross-sections of photonuclear reactions will be conducted using the photon difference method. In this case, experimental studies are planned on an output device where a deflecting magnet is used to purify the bremsstrahlung $\gamma$-flux from electrons. This experimental configuration allows for improved parameters of the bremsstrahlung spectrum on the target, and its modeling can be conducted with greater statistics due to the absence of an aluminum absorber in the electron path.

Experiments will be performed using isotopic target samples with a high degree of enrichment (Mo, Sb, Ag, etc.). This will provide different information about the average cross-sections of photonuclear reactions, in contrast to results from natural targets. Additionally, differences in isomeric ratios IR($E_{\gamma max}$) will be investigated for the same final isomeric pair, however when excited from different initial nuclei.

### 3.3 RESEARCH ON EFFECT OF RADIATION ON PROPERTIES MATERIALS

Materials and devices used in accelerator technology and high-energy physics detectors are subject to significant radiation loads. In this regard, when developing new installations, much attention is paid to conducting appropriate studies on radiation resistance.

Further development of accelerator technology is associated with the search for new methods charged particle beams acceleration. One of the directions of these studies is based on the use of dielectric materials. This applies both to acceleration methods in wake fields, excited by electron bunches, and to new dielectric assist accelerating structures with the traditional RF power input. When studying the effect of radiation on dielectric characteristics in the microwave range, it is important to measure the permittivity and dielectric losses directly during irradiation. We have developed a corresponding technique based on the use of open resonators and have measured some types of microwave ceramics [40].

Together with CERN, ISMA of the NAS of Ukraine and Taras Shevchenko National University of Kyiv a series of experiments were conducted to study the radiation resistance of elements PLUME detector LHCb experiment. The optical elements (quartz tablets for Cherenkov light generation, plastics scintillators of various types, quartz fibers for light transmission, several optical contact materials etc.) were irradiated on the LUE-40 with doses close to the operating conditions of this elements - absorbed dose $(2-8)10^5$ Gy and neutron fluence $10^{14}$ n/cm$^2$ [41-43].

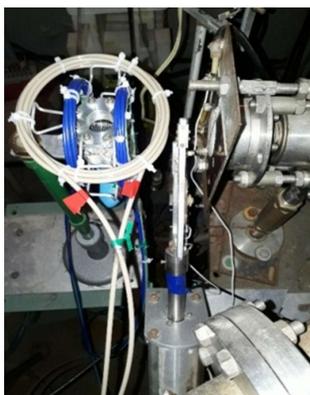

*Fig.5 Irradiation of quartz fibers for light transmission*

New types of plastic scintillators based on polystyrene are being developed at the ISMA [44]. Plastic scintillators are widely used in high-energy physics experiments and must have sufficient radiation hardness. To study effect radiation on optical and scintillation properties of the scintillators they are irradiated on the LUE-40 with an electron beam. Typical beam parameters are energy of 40-42 MeV and average current up to 5 μA. According to experimental conditions, the absorbed dose varies from 50 to 400 kGy with absorbed dose rate of 90-120 Gy/s.

### CONCLUSIONS

Currently the operability of all systems of the LUE-40 or has been restored, the beam parameters correspond to the passport ones. The accelerator is the only operating source of electrons with an energy of up to 100 MeV in our country. The capabilities of the LUE-40 for the formation and diagnostics of an electron beam, deep regulation of its energy, as well as the creation of output devices for generating secondary bremsstrahlung and neutron radiation provided conditions for conducting experimental nuclear physics research. In the near future, it is planned to continue research on multi-particle photonuclear reactions on atomic nuclei of medium and heavy masses in the region of boundary bremsstrahlung energies of γ-quanta $E_{\gamma max}$ up to 100 MeV. Currently at the LUE-40, the studies are being carried out aiming at improving the accuracy of the cross-section data for photonuclear production of a number of medical isotope and optimization of such processes using the developed techniques. Experiments are ongoing to study the effects of radiation on the properties of scintillators and other polymeric materials.

### ACKNOWLEDGMENT

The authors express their gratitude to the staff of the NDK "Accelerator", who ensured the operability of all accelerator systems and made it possible to carry out the experiments described in the article.

## УСТАНОВКА НА БАЗІ ЛІНІЙНОГО ПРИКОРЮВАЧА ЕЛЕКТРОНІВ LUE-40 – СУЧАСНИЙ СТАН ТА ОСНОВНІ НАПРЯМКИ ДОСЛІДЖЕНЬ

**М.І. Айзацький, В.М. Борискін, І.О. Чертищев, О.С. Деєв, К.Ю. Крамаренко, В.А. Кушнір, В.В. Митроченко, С.М. Олійник, С.О. Пережогін, С.М Потін, Л.І. Селіванов, І.С. Тімченко, Ю.О. Титаренко, В.Ю. Титов, В.Л. Уваров**

Описано стан установки на базі лінійного прискорювача електронів ЛУЕ-40 після реставрації у 2023 році. Розроблено та створено додаткові пристрої для проведення ядерно-фізичних досліджень. Наведено параметри пучка. Діапазон енергій електронів розширено і нині становить 16–95 МеВ при середньому струмі пучка до 6 мкА. Описано основні напрями наукових досліджень та наведено отримані результати